\documentclass[12pt]{iopart}

\usepackage{graphicx, epsfig, subfigure}

\begin{document}
 
\title{ A New Approach to the Yang-Mills Gauge Theory of Gravity and It's Applications } 

\author{Yi Yang}
\address{The Ohio State University, Columbus, OH, USA}
\ead{yi.yang@cern.ch}
\author{Wai Bong Yeung}
\address{Institute of Physics, Academia Sinica, Taipei, Taiwan, ROC}
\ead{phwyeung@phys.sinica.edu.tw}

\begin{abstract}
We shall give dynamics to our spacetime manifold by first identifying the local affine symmetry as the characterizing symmetry for our geometry $\acute{a}$la Felix Klein, this symmetry is imposed on us by the Law of Inertia and the Law of Causality. 
We then prescribe 16 gauge vector bosons to this symmetry $\acute{a}$la Yang and Mills. 
The locally affine symmetric Yang-Mills Lagrangian in the presence of a background world metric, and the corresponding equations of motion, are respectively constructed and derived. 
Spontaneous breaking of the local affine symmetry to the local Lorentz symmetry is achieved by classical solutions to the equations of motion. 
In these classical solutions, the 16 gauge vector bosons are shown to select the Schwarzschild metric as one among the admissible background world metrics. 
Classical gravity is thus be expressed by a spontaneously broken Erlangen program. 
We shall also show that this Yang-Mills gauge theory of gravity can give an explanation of the form of the galactic rotation curves, of the amount of intergalactic gravitational lensing, and of the accelerating expansion of the Universe.
\end{abstract}

\maketitle

\section{Introduction}
The four fundamental interactions in physics are described by two different disciplines. 
The gravitational interaction follows the curved spacetime approach of the General Theory of Relativity laid down by Einstein in which the dynamical variable is the metric tensor, while the electroweak and strong interactions follow the local gauge vector boson approach pioneered by Yang and Mills in which the dynamical variables are the vector bosons. 
Both disciplines give spectacular success in terms of experiments and physical observations, despite of the fact that they look very different. 

There are, by now, many research works done in trying to put these two disciplines into one single footing. 
Some people try to visualize gauge vector boson interactions as geometrical manifestations in a higher dimensional manifold with our spacetime as a four dimensional sub-manifold~\cite{Ref:1, Ref:2}. 
Other people try to consider the geometrical gravitation theory in the form of a local gauge theory~\cite{Ref:3, Ref:4}. 
All of these ideas are met with difficulties in one way or the other.   

And in the past decades many new discoveries in astronomy might imply that the General Theory of Relativity is not sophisticated enough to explain the new phenomena.   
For example, stellar objects at the spiral arms of galaxies are orbiting at faster speeds than that can be explained by the Keplerian motions. 
To overcome this difficulty, people assume that some extra matter, not visible to us, is giving an extra pull on these stellar objects. 
Extra light deflections, as observed in the intergalactic gravitational lensing, are also ascribed to the existence of this extra matter. 
This is known as the $Dark$ $Matter$ problem. 
An equally well known fact is that the Universe is accelerating in its expansion. 
This is in contrast to our expectation that the Universe should be decelerating, unless some extra energy is kept pumping into the Universe. 
This is the so called the $Dark$ $Energy$ problem

In this article, we shall give a new approach to use the Yang-Mills method to construct a local Gauge Theory of Gravity, and shall show that this theory can successfully solve the $Dark$ $Matter$ and $Dark$ $Energy$ problems.  

\section{A Spacetime Manifold with a Non-Dynamical Background World Metric}
Here we shall consider the local gauge theory approach of gravitation, albeit in a new context. 
A non-Minkowskian world metric for our spacetime is always regarded as what makes it curved. 
However, it is very difficult to regard the world metric (or more precisely its corresponding vierbein fields) as gauge fields because of the peculiar way it appears in the action that determines physics.

In the following we shall assign the world metric of our spacetime with a much limited role. 
We shall assume that the only function of the world metric $g_{\mu\nu}$ is to give us world distance (and hence world volume element $\sqrt{-g}d^4x$), and will have no dynamical terms (terms that contain spatial or temporal derivatives of $g_{\mu\nu}$) in the action. 
This assumption frees us from taking the global Minkowskian metric as the de facto world metric and sets the notion that no particular metric is a priori world metric for physics. 
In this sense, the world metric for our spacetime serves just as an arbitrary background of measuring clock and stick in our discussion of physics.

With a given world metric $g_{\mu\nu}(x)$ at the point $x$ with world coordinates $x^{\mu}$, a set of vierbein fields $e^{a}_{\ \mu}(x)$ will follow. 
These vierbein fields are defined in a locally flat patch that is assumed to be equipped with an arbitrary but given local Minkowskian frame whose coordinates and metric are $x^a$ and $\eta_{ab}$ respectively. 
The differentials of these two coordinate systems define the vierbein fields as $dx^a = e^a_{\ \lambda}dx^{\lambda}$, and hence the world metric and the vierbein fields will then be related by 
\begin{equation}
\label{eq:vierbein}
\eta_{ab} e^a_{\ \mu} e^b_{\ \nu} = g_{\mu\nu}.
\end{equation}
Here, and in the following, the Latin indices will signify the Minkowskian components while the Greek indices will mean the world ones.

\section{Some Basic Physical Principles that are Invariant Under the Proposed Local Coordinate Transformations }
On a locally flat patch around a point of our spacetime is where we do our physics. 
Even though we have already had a local Minkowskian system $x^a$ and $\eta_{ab}$ on that patch, we may still have the freedom to re-label the points on that patch with different local coordinate systems, for example, by rotating and stretching these local Minkowskian coordinate axes. 
These rotations and stretches are reflections of the relative physical situations experienced by local observers. 
For examples, a relative rotation of two local Minkowskian frames is always regared as a relative Lorentzian motion between the two local observers. 

The form of the admissible local coordinate transformations depends on what are the physical principles that, we hope, to remain invariant under these coordinate changes.

Here, we believe that the law of inertia should remain intact under these expected coordinate transformations. 
This means that the concept of straight lines should be preserved, as an object moving in straight line in one coordinate system should remain moving in straight line in another coordinate system. 
Also light should propagate in straight lines in whatever coordinate system we are using. 
Causality is also a very important concept in physics, and hence the order of points and the ratio of segment lengths in a straight line should not change with a change in coordinate system. 
And of course, the concept of parallelness should also be preserved because two parallel moving objects, as well as parallel light rays, should remain parallel under a change of coordinate system.

Those transformations which maintain collinearity, order of points and invariant segment ratios in straight lines, and parallelness are, in fact, the affine transformations of Euler~\cite{Ref:5}. 
Affine transformations are sometimes grouped together as dilations, rotations, shears and reflections. 
We shall call collectively transformations that are not rotations as strains.

\section{Marriage of the Erlangen Program with the Yang-Mills Doctrine -- a Way to Give Dynamics to a Geometry}
Here we want to emphasize that our choice of the affine transformations as our admissible local coordinate transformations comes from physics. 
It comes from our belief that these admissible transformations should leave the above said physical principles invariant. 
And if we are going to call such a chosen local coordinate system as a chosen local geometric setting, then we can say that physics is assumed to be invariant under a change of local geometric setting. 
In physical language, what his means, is that physics looks the same to all observers having the proposed relative physical situations. 
And this can be regarded as an extension of the Principle of Relativity of the Special Theory of Relativity to a general spacetime manifold. 

These transformations form a Lie group, called the local affine group. 
It was Felix Klein who first suggested of classifying geometries by their underlying symmetry groups, starting with the Projective Geometry (our affine geometry is a restriction of the Projective Geometry). 
Such a mathematical program is called the Erlangen Program~\cite{Ref:6}.

Since matter, which are world objects, are described by local fields with the reference to a local coordinate system. 
These local fields could have structures that depend on the geometric setting chosen at that point. 
For example, if we want to describe the physics of an electron, it may be convenient to choose the local affine rotations as our local Lorentz transformations. Then an affine rotated setting will give a set of Lorentz transformed fields.

As we believe that the relative differences of the local fields of the same world object at two space-time points arising from different geometric settings are physically meaningless, we have to find some way to counteract such variations. 
Similar to what have been done by Yang and Mills~\cite{Ref:7}, we shall introduce a set of vector bosons to do these counteractions. 

Note that the introduction of vector bosons, as suggested by Yang and Mills, were originally used to facilitate the local identifications of internal quantum numbers for quantum systems. 
Here we extend their ideas to the local identifications of geometric settings in our spacetime.

At any point of our spacetime, these vector bosons can be transformed away locally by a suitable choice of the coordinate system at that point. 
A more familiar way of saying this is that these vector fields are locally equivalent to a transformation of the coordinate system. 
One can then draw the strong analogy between the above statement and the Principle of Equivalence of Einstein which states that "there is a complete physical equivalence of a gravitational field and a corresponding acceleration of the reference system". The Principle of Equivalence is the founding principle taken by Einstein to build his General Theory of Relativity, and the gauge principle stated in the above will be shown, as given below, to give a theory of vector gravity. 

In the following, these vector boson fields are regarded as dynamical variables. 
Their dynamics will be fabricated so as to ensure that physics be invariant under local affine transformations. 
And this will be done, again by following Yang and Mills, by first constructing the Lagrangian that is locally affine symmetric.

\section{The Affine Group $GL(4\ R)$}
For a four dimensional patch, these affine transformations can be carried out by $4\times4$ invertible real matrices, either actively or passively. 
All these $4\times4$ invertible real matrices form a Lie group called the Real General Linear group of dimension four and is designated as $GL(4\ R)$. 
Hence the $GL(4\ R)$ will be synonymous with our affine group. 
The $GL(4\ R)$ has two sets of generators. 
The six anti-symmetric generators $J^{ab}$ generate the rotations while the ten symmetric generators $T^{ab}$ generate the strains. 
They satisfy the following commutation relations~\cite{Ref:8}   
\begin{eqnarray}
\label{eq:commu_rel}
      \left[J^{ab}, J^{cd}\right] &=& -i \{ \eta^{ac}J^{bd} - \eta^{ad}J^{bc} - \eta^{bc}J^{ad} + \eta^{bd}J^{ac} \}; \nonumber \\
      \left[J^{ab}, T^{cd}\right] &=& -i \{ \eta^{ac}T^{bd} + \eta^{ad}T^{bc} - \eta^{bc}T^{ad} - \eta^{bd}T^{ac} \};   \\
      \left[T^{ab}, T^{cd}\right] &=& i \{ \eta^{ac}J^{bd} + \eta^{ad}J^{bc} + \eta^{bc}J^{ad} + \eta^{bd}J^{ac} \}. \nonumber 
\end{eqnarray}
These generators, when combined together as $M^{ab}=\frac{1}{2}(T^{ab} + J^{ab})$, and with the indices lowered by $\eta_{ab}$, give a compact commutation relation of the form of
\begin{eqnarray}
\label{eq:commu_gen}
   [M_a^{\ b}, M_c^{\ d}] = i \delta^{b}_c M_a^{\ d} - i \delta^{d}_a M_c^{\ b}
\end{eqnarray}
We know that the $GL(4\ R)$ with the defining Lie Algebra given in Eq.~\ref{eq:commu_gen} has no presupposition of the existence of the Minkowskian metric on the locally flat patch of our spacetime. 
We are introducing $GL(4\ R)$ into physics in our way because we want to emphasize that there are some very fundamental laws, namely, the Law of Inertia and the Law of Causality, working together to impose the $GL(4\ R)$ symmetry onto our spacetime. 
"Invariance dictates interaction", said C. N. Yang.
This doctrine seems working well in the electroweak and the strong interactions. 
We propose in this article that this doctrine could apply to gravity too.

\section{ The Yang-Mills Action for the Local $GL(4\ R)$ in the Presence of a Background World Metric }
The Yang-Mills gauge potentials for the $GL(4\ R)$ are
\begin{equation}
\label{eq:YM_pot}
   A_{\mu} = A^{m}_{\ n\mu}M_{m}^{\ n}.
\end{equation}
Note that there are totally sixteen gauge bosons $A^{m}_{\ n\mu}$ appearing in our theory. 
The antisymmetric parts of $A^{m}_{\ n\mu}$ go with the generators $J_{m}^{\ n}$ while the symmetric parts go with the generators $T_{m}^{\ n}$. 
These sixteen gauge bosons are world vector fields.

The Yang-Mills field strength tensor $F_{\mu\nu}$ is
\begin{eqnarray}
\label{eq:f_munu}
 F_{\mu\nu} &=&  \partial_{\mu}A_{\nu} - \partial_{\nu}A_{\mu} - i\left[A_{\mu}, A_{\nu}\right] \nonumber \\
            &=& ({\partial_{\mu}A^{m}_{\ n\nu} - \partial_{\nu}A^{m}_{\ n\mu} + A^{m}_{\ p\mu}A^{p}_{\ n\nu}- A^{m}_{\ p\nu}A^{p}_{\ n\mu} } ) M_{m}^{\ n} \nonumber \\
            &\equiv& F^{m}_{\ n\mu\nu}M_{m}^{\ n}. 
\end{eqnarray}

The Yang-Mills Lagrangian, which is invariant under the local $GL(4\ R)$ transformations, is 
\begin{equation}
\label{eq:YM_lag}
  \mathcal{L}_{\rm YM} = \frac{1}{2}{\rm Tr}F_{\mu\nu}F^{\mu\nu}.  
\end{equation}
It is interesting to see how we can evaluate the trace of the products of the generators of $GL(4\ R)$. 
For $GL(4\ R)$, there exists a relation between the trace of the products and the product of the traces of the generators and the bilinear Killing form of the Special Linear group $SL(4\ R)$ (as denoted by a tilde), namely 
\begin{equation}
  \label{eq:gen_trace}
  {\rm Tr}(M_{a}^{\ \ b} M_{c}^{\ \ d}) = \frac{1}{8}K(\tilde{M}_{a}^{\ \ b} \tilde{M}_{c}^{\ \ d}) + \frac{1}{4}{\rm Tr}(M_{a}^{\ \ b}){\rm Tr}(M_{c}^{\ \ d}),
\end{equation}
where
\begin{equation}
  \label{eq:killing_form}
  K(\tilde{M}_{a}^{\ \ b}\tilde{M}_{c}^{\ \ d}) = {\rm Tr}(ad(\tilde{M}_{a}^{\ \ b})ad(\tilde{M}_{c}^{\ \ d})),
\end{equation}
$K$ is the bilinear Killing form and $ad(\tilde{M}_{a}^{\ \ b})$ is the Adjoint Representation of $\tilde{M}_{a}^{\ \ b}$ which is traceless. 
The calculation of the trace of the products of the Adjoint Representation is direct, and it is 
\begin{equation}
  \label{eq:gen_trace_adj}
  {\rm Tr}(ad(\tilde{M}_{a}^{\ \ b})ad(\tilde{M}_{c}^{\ \ d})) = 8 \delta_{a}^{d}\delta_{c}^{b} - 2 \delta_{a}^{b}\delta_{c}^{d}.
\end{equation}
The calculation of ${\rm Tr}(M_{a}^{\ \ b})$ is easy, too. 
The generators of $GL(4\ R)$ differs from the generators of $SL(4\ R)$ by a dilatation part of ${\rm T}_{a}^{\ \ b}$, which is $\frac{1}{4}\delta_{a}^{b} {\rm I}$.
In other words, ${\rm T}_{a}^{\ \ b} = \tilde{\rm T}_{a}^{\ \ b} + \frac{1}{2}\delta_{a}^{b}{\rm I}$, and hence we have 
\begin{equation}
  \label{eq:gen_trace_2}
    {\rm Tr}(M_{a}^{\ \ b}) = \delta_{a}^{b}. 
\end{equation}
Combing Eq.~\ref{eq:gen_trace}, Eq.~\ref{eq:killing_form}, Eq.~\ref{eq:gen_trace_adj} and Eq.~\ref{eq:gen_trace_2} gives 
\begin{equation}
  \label{eq:gen_trace_3}
    {\rm Tr}(M_{a}^{\ \ b} M_{c}^{\ \ d}) = \delta_{a}^{d}\delta_{c}^{b}. 
\end{equation}
This is independent of the kind of representation chosen for $GL(4\ R)$. 
Note that the result given in Eq.~\ref{eq:gen_trace_3} comes not only from the structure constants of Eq.~\ref{eq:commu_gen}, but also comes from a judicious choice of the dilation part of the $GL(4\ R)$ generators. 

And hence the affine symmetric Yang-Mills action $\rm S_{YM}$, in the presence of the background world metric $g_{\mu\nu}$, will be
\begin{equation}
\label{eq:YM_action}
  {\rm S_{YM}}\left[g, A, \partial A \right] = \kappa \int \sqrt{-g} d^4x g^{\mu\mu'} g^{\nu\nu'} ( \delta_{a}^{d}\delta_{c}^{b} ) F^a_{\ b\mu\nu} F^c_{\ d\mu'\nu'},
\end{equation}
$\kappa$ is a dimensionless coupling constant of the theory. 

Of course, the total action will also contain a piece coming from the matter fields which are supposed to couple gauge invariantly to the gauge fields~\cite{Ref:4, Ref:8}, and thus making the total action $\rm S_{total}$ as 
\begin{equation}
\label{eq:tot_action}
{\rm S_{total} } =  {\rm S_{YM}} + \int \sqrt{-g}d^4x \mathcal{L}_{\rm matter}.
\end{equation}

\section{How the sixteen Gauge Vector Bosons Select the Background World Metric in Classical Physics }
Even though the world metric serves just as a background measuring clock and stick for our spacetime, it does contribute to the Feynman amplitudes in calculating physical processes because
\begin{equation}
\label{eq:amp}
  {\rm Amplitude} = \int e^{i \frac{ S_{\rm total} }{\hbar} } {\mathcal D}\left[g_{\mu\nu}, A^{m}_{\ n\mu}, \rm{matter \ fields}\right], 
\end{equation}
where ${\mathcal D}\left[g_{\mu\nu}, A^{m}_{\ n\mu}, \rm{matter \ fields}\right]$ is the $GL(4\ R)$ invariant measure over the fields. 

In the following, we shall study the $classical$ physics implied by Eq.~\ref{eq:amp}. 
And this can be done by looking at the extrema of the action. 

The role played by $g_{\mu\nu}$ in classical physics would be clear if we were able to integrate out the $GL(4\ R)$ gauge field and then solve for the equation of motion containing only the metric. Such similar programs were indeed being tried many times before and were never made to work. 

Here we want to suggest a novel way to attack this classical problem by solving the classical Yang-Mills equation in a background world metric. 
Solving a classical Yang-Mills gauge equation in a background world metric is by no means easy. 
Fortunately, we are able to show that the $GL(4\ R)$ gauge theory is identical to a geometric theory quadratic in the Riemannian tensor albeit that the metric and connections are now independent variables. And thanks to the works of many people, this geometric theory was shown to contain the Schwarzschild metric and some other world metric as solutions. In the following, we are trying to show how this can happen. 

A particular choice of the background world metric, and a particular set of the affine fields and matter fields, that together extremize the total action, will give us the physics that we are observing in the classical world. 
In other words, any choice of the world metric is allowed as background metric, but only those metrics that satisfy the extremal conditions are what we are experiencing classically. 
These extremal conditions are
\begin{eqnarray}
\label{eq:var_action}
  \left. \frac{\delta S_{\rm total}}{\delta g_{\theta\tau}} \right \vert_{A} &=& \sqrt{-g}(F^{a}_{\ c\theta\rho}F^{c \ \ \rho}_{\ a\tau} - \frac{1}{4}g_{\theta\tau}F^{a}_{\ c\xi\rho}F^{c \ \xi\rho}_{\ a} - \frac{1}{4\kappa} T_{\theta\tau} ) = 0 ;\nonumber \\
  \left. \frac{\delta S_{\rm total}}{\delta A^{m}_{\ n\nu}} \right \vert_{g} &=& D_{\rho}(A)(\sqrt{-g}F^{n\ \rho\nu}_{\ m}) - \frac{1}{\kappa}\sqrt{-g}S^{n\ \nu}_{\ m} = 0 ;\\
  \frac{\delta S_{\rm total}}{\delta \rm{matter\ fields}} &=& 0 .\nonumber
\end{eqnarray}
where $D_{\rho}(A)$ denotes the Yang-Mills gauge covariant differentiation. 
The $T_{\theta\tau}$ and $S^{n\ \nu}_{\ m}$ are respectively the metric energy-momentum tensor and gauge current tensor of the source matter. 

This is the set of equations that we are proposing to describe the spatial and temporal evolution of the classical $GL(4\ R)$ Yang-Mills and the matter fields, in our spacetime which has a background world metric $g_{\mu\nu}$.

It turns out that not all world metrics can sustain a classical $GL(4\ R)$ Yang-Mills field, only some selected ones can do. 
Putting Eq.~\ref{eq:var_action} in words: the solved $A^{m}_{\ n\nu}$ from the second equation (which is the Yang-Mills equation) will be functionals of $g_{\mu\nu}$. 
And when we plug the solved $A^{m}_{\ n\nu}$ into the first equation, it will become an equation for $g_{\mu\nu}$. 
And from this first equation we shall select the world metrics for our classical world. 

Up to this point, we have not used any sophisticated concepts in geometry such as the connection, the parallelism and the curvature. 
The only thing we have used that may have something to do with geometry is that our spacetime should have a metric, telling us how to measure distance and volume. 
So the geometry of our spacetime is not Riemannian, and is not even affine; it is just metrical.

From our point of view, a metric is fundamental and is needed if we want to construct an action from some local fields. 
The concepts of connection, parallism and curvature are, however, not fundamental. 
If we can show that the Schwarzschild metric and some other metric can follow from our Eq.~\ref{eq:var_action}, then we can claim that the description of gravity needs no sophisticated geometric ideas other than that of a metric. 

Before we embark on further discussions, we should clarify the role played by our dimensionless parameter $\kappa$ and should also clarify the way how the Newtonian gravitational constant $G$ makes its appearance in gravitation phenomena. 

$\kappa$ appears because there is an arbitrariness in fixing the relative scale between the matter and gauge parts of the Lagrangian. 
The value of $\kappa$ should have no significance in physics, and can be absorbed into the definition of the matter field. 

What have significance on physics are the integration constants that accompany the solutions to the equation of motion. 
Because $g_{\mu\nu}(x)$ is dimensionless, $x$ must be scaled by some integration constant $l$ which has the dimension of length. 
It is a different $l$ that gives a different strength in gravitational interaction. 
The fact that $l$ is proportional to the inertial mass $M$ of the gravitational source, $l = 2GM$, is inferred by comparing the weak field limit of the solution with the Newtonian gravitational potential. 
This situation also happens in General Relativity. 

\section{How the Yang-Mills Equation Becomes the Gravity Equation}
If we have the experience in solving Eq.~\ref{eq:var_action} in its gauge form, we would have the above claim that classical gravity comes from the Yang-Mills gauge theory of $GL(4\ R)$ done. 
Just as we have said before, solving the $GL(4\ R)$ Yang-Mills equations in a background world metric is by no means easy. 
One way that could help us in solving Eq.~\ref{eq:var_action} is to cast this equation into a form that we are acquainted with. 
From the Yang-Mills fields $A^{m}_{\ n\nu}$ and the vierbein fields $e^{a}_{\ \mu}$ we can construct some new fields called the $\Gamma$ fields by~\cite{Ref:9}
\begin{equation}
\label{eq:YM_pot_new}
    A^{m}_{\ n\mu} = e^{m}_{\ \rho} e_{n}^{\ \tau} \Gamma^{\rho}_{\tau\mu} + e^{m}_{\ \tau} \partial_{\mu} e_{n}^{\ \tau}, 
\end{equation}
and then substitute the sixty-four variables $A^{m}_{\ n\mu}$ by the new sixty-four variables $\Gamma^{\rho}_{\tau\mu}$, by plugging $A^{m}_{\ n\mu}$ into $F^{m}_{\ n\mu\nu}$ in Eq.~\ref{eq:f_munu}. 
Note that $\Gamma^{\rho}_{\tau\mu}$ is defined by $A^{m}_{\ n\mu}$ and has, so far, nothing to do with connections. 
Miraculously, the Yang-Mills field strength tensor can be re-expressed in the $\Gamma$ fields in a very simple way as~\cite{Ref:10}
\begin{eqnarray}
\label{eq:YM_f_munu}
    F^{m}_{\ n\mu\nu} &=& e^{m}_{\ \lambda} e_{n}^{\ \sigma} ( \partial_{\mu}\Gamma^{\lambda}_{\sigma \nu } - \partial_{\nu}\Gamma^{\lambda}_{\sigma \mu } + \Gamma^{\lambda}_{\kappa\mu} \Gamma^{\kappa}_{\sigma\nu} - \Gamma^{\lambda}_{\kappa\nu} \Gamma^{\kappa}_{\sigma\mu} ) \nonumber \\
    &\equiv& e^{m}_{\ \lambda} e_{n}^{\ \sigma} R^{\lambda}_{\ \sigma \mu\nu}, 
\end{eqnarray}
where we have used $R^{\lambda}_{\ \sigma\mu\nu}$ to stand for $( \partial_{\mu}\Gamma^{\lambda}_{\sigma \nu } - \partial_{\nu}\Gamma^{\lambda}_{\sigma \mu } + \Gamma^{\lambda}_{\kappa\mu} \Gamma^{\kappa}_{\sigma\nu} - \Gamma^{\lambda}_{\kappa\nu} \Gamma^{\kappa}_{\sigma\mu} )$.    
Note that our $R^{\lambda}_{\ \sigma\mu\nu}$, though looks like the Riemann curvature tensor, is, in fact, a derived quantity coming from the Yang-Mills field tensor. 
Plugging the result in Eq.~\ref{eq:YM_f_munu} into the Yang-Mills action in Eq.~\ref{eq:YM_action}, the Yang-Mills action will then look like
\begin{equation}
\label{eq:YM_action_new}
  {\rm S_{YM}}\left[g, A, \partial A \right] 
  = {\rm S_{YM}}\left[g, \Gamma, \partial\Gamma \right] 
  = \kappa \int \sqrt{-g} d^4x g^{\mu\mu'} g^{\nu\nu'} ( R^{\lambda}_{\ \sigma\mu\nu}R^{\sigma}_{\ \lambda\mu'\nu'} ). 
\end{equation}
Note that there is only one term quadratic in the Riemann curvature appearing in the action. 
When the Yang-Mills action $\rm S_{YM}$ is expressed in terms of the new fields, the Yang-Mills equation can, of course, be obtained by extremizing $\rm S_{YM}[g, \Gamma]$ with respect to $\Gamma^{\rho}_{\tau\mu}$. 
Hence we have arrived at the very important fact that the Yang-Mills Equation for the affine symmetry group can be written as
\begin{equation}
\label{eq:YM_act_var}
    \left. \frac{\delta \rm S_{YM}[g, \Gamma]}{\delta \Gamma^{\rho}_{\tau\mu}}\right \vert_{g} + \left. \frac{\delta \rm S_{matter}}{\delta \Gamma^{\rho}_{\tau\mu}}\right \vert_{g} = 0.
\end{equation}
Variation with the $g_{\mu\nu}$ will give us another equation, which we shall display as           
\begin{equation}
\label{eq:tot_act_var}
   \left. \frac{\delta \rm S_{total}}{\delta g_{\theta\tau}}\right \vert_{\Gamma} =  \left. \frac{\delta \rm S_{total}}{\delta g_{\theta\tau}}\right \vert_{A} +  \left. \frac{\delta \rm S_{total}}{\delta A}\right \vert_{g}  \left. \frac{\delta A}{\delta g_{\theta\tau}}\right \vert_{\Gamma} = 0.
\end{equation}

Note that Eq.~\ref{eq:tot_act_var} is in fact a combined result of the first equation and the second equation of Eq.~\ref{eq:var_action}, because the action can depend on $g_{\mu\nu}$ through its dependence on $A^{m}_{\ n\nu}$ when $\Gamma^{\rho}_{\tau\mu}$ is held fixed. 
One interesting thing that is worth noting is that if we are going to integrate over all the distinctive $\Gamma^{\rho}_{\tau\mu}$ (apart from a diffeomorphism of the entire spacetime), we are in fact factoring out the group volume of $GL(4\ R)$ as was shown explicitly in Eq.~\ref{eq:YM_pot_new}. 
The legitimate measure over the fields will then be ${\mathcal D}\left[g_{\mu\nu}, \Gamma^{\rho}_{\tau\mu}, \rm{matter \ fields}\right]$. 

Now Eq.~\ref{eq:YM_pot_new} looks like the famous geometric relation between the affine connections and the spin connections under the tetrad postulate, if we regard $\Gamma^{\rho}_{\tau\mu}$ as the affine connections and $A^{m}_{\ n\nu}$ as the spin connections. 
And Eq.~\ref{eq:YM_action_new}, Eq.~\ref{eq:YM_act_var} and Eq.~\ref{eq:tot_act_var}, together, look like a geometric theory with a geometric Lagrangian and a set of geometric equations of motion that we have encountered frequently in talking about gravity. 
Hence it is natural for us to put all these in the following geometric jargons: that a parallel connection $\Gamma^{\rho}_{\tau\mu}$ is introduced into the spacetime, that a Riemann curvature tensor is constructed, that a gravitational Lagrangian is formed out of the product of the Riemannian curvature tensor, that we are trying to obtain the gravity equations by varying the connections and the metric independently $\acute{a}$la Palatini~\cite{Ref:11}, and that gravity is a kind of Metric-Affine Theory~\cite{Ref:4}.
From the above discussions, it is now clear that what all these things that were done in the past, were done in pieces by following the doctrines laid down by Klein, Einstein, Yang and Mills and Feynman, either knowingly or unknowingly.

This form of the gravitational action of Eq.~\ref{eq:YM_action_new} has presented itself many times in the history of the development of the geometric theory of gravity, but is, in fact, carrying very different information at each one of the presentations. 
The point of focus is on the relation between the metric and the connections.

In its very early version, as proposed by Hermann Weyl~\cite{ref:old_1}, the connections that appear in the theory are nothing but the Christoffel symbols which are of first derivatives in the metric. 
This will result into a theory in which the metric is the only dynamical variable, and the variation with respect to the metric will give an equation of motion of higher order derivatives. 
It is well known that such a theory will possess runaway solutions.
     
Later Yang~\cite{ref:old_2}, also regarded the connections as the Christoffel symbols at the start, but varied the connections instead in order to get the equation of motion. 
The final result is, again, an equation of higher derivatives in the metric.
     
Stephenson~\cite{ref:old_3}, put the anti-symmetric parts of the connections equal to zero, and regarded the symmetric parts of the connections and the metric as independent variables. 
And he obtained two equations of motion by varying both the metric and symmetric parts of the connections independently.
      
On the other hand, some people identify the symmetric parts of the connections as the Christoffel symbols and regard the anti-symmetric parts and the metric as independent variables. 
Those people working on the Poincare Gauge Theory of Gravity are taking this point of view~\cite{ref:old_4}.
       
For us, the full connections, both the symmetric and the anti-symmetric parts, as well as the metric are independent variables. 
In this theory of gravity, the connections are just the transformed $GL(4\ R)$ Yang-Mills vector potentials $A^{m}_{n\mu}$, and can be taken as being independent of $g_{\mu\nu}$.

Here we feel obligatory for us to re-assert the reason that we are making excursion into the land of geometry is because we want to make use of some of the results known to the people working in the geometric theory of gravity. And of course, we also want to know how our proposed theory looks like in geometrical languages. 

The different choices of the content coded in the Riemann curvature tensor give different stories for physics. 
For example, for the Weyl theory, the metric is the only dynamical variable, and hence the action will contain kinetic terms that have derivatives that are of orders higher than two. 
And when we look for the possible propagation modes in the theory, which can be obtained by looking at the inverse of the kinetic term, we will find that there will be propagators having the wrong signs, which will correspond to unphysical states called the ghosts or tachyons, and will end up into an unstable theory with the so called Ostrogradski instability~\cite{ref:old_6}.
     
For us, the metric is a non-dynamical background field~\cite{ref:old_5}, and the only dynamical variables are the connections which obey an equation second order in space and time derivatives. 
The propagating modes are the 16 vector bosons and nothing else. 
And hence our theory will contain no ghost and no Ostrogradski instability. 
We are not having the pathologies that are affecting quadratic curvature theories in which the metric is dynamical and is compatible with the connections. 

\section{The Schwarzschild and the TPPN Metrics are Induced by the sixteen Gauge Vector Bosons}
Let us now concentrate ourselves on the ``vacuum" solutions of Eq.~\ref{eq:YM_action_new}, Eq.~\ref{eq:YM_act_var} and Eq.~\ref{eq:tot_act_var}. 
By ``vacuum" here we shall mean the case where all the matter fields are absent, except possibly at the source point. 
However, the $GL(4\ R)$ gauge vector fields may not necessarily be vanishing. 

A trivial solution with a global Minkowskian world metric and vanishing $GL(4\ R)$ gauge potentials can be inferred immediately from the equations. 

Solving the ``vacuum'' $GL(4\ R)$ Yang-Mills equation in the presence of a background metric will be prohibitively difficult, even in the spherically symmetric cases (see Appendix).
However, we are fortunate enough to have shown that the $GL(4\ R)$ Yang-Mills gauge theory in a background metric is equivalent to the metric-affine theory of gravity with a Lagrangian quadratic in the Riemann curvatures, namely Eq.~\ref{eq:YM_action_new}. 
In this geometric language, we can then invoke the concept of metric-affine compatibility and of the concept of torsion. 
It is under the requirements of compatibility and torsionless that we are trying to see if we can find a subclass of solutions to Eq.~\ref{eq:YM_action_new}, Eq.~\ref{eq:YM_act_var} and Eq.~\ref{eq:tot_act_var}.

In the following, we shall search for solutions $under$ the ansatz that the world metric $g_{\mu\nu}$ and the connections $\Gamma^{\theta}_{\tau\xi}$ are compatible with each other. 
We shall call this ansatz the Compatibilty Ansatz (CA). 
And we consider only solutions that are torsionless. 
Then the ``vacuum" version for Eq.~\ref{eq:YM_action_new}, Eq.~\ref{eq:YM_act_var} and Eq.~\ref{eq:tot_act_var} can be written as
\begin{eqnarray}
\label{eq:ca_sky}
    && \left[ \frac{\delta}{\delta \Gamma^{\theta}_{\tau\xi}} \int \sqrt{-g} d^4x g^{\mu\mu'} g^{\nu\nu'} (  R^{\lambda}_{\ \sigma\mu\nu}R^{\sigma}_{\ \lambda\mu'\nu'} ) \right]_{{\rm CA \ + \ torsionless}} = 0, \nonumber \\  
\end{eqnarray}
which means 
$\nabla_{\rho}(R_{\theta}^{\ \tau\rho\xi}) = 0$, and
\begin{eqnarray}
\label{eq:ca_step}
   H^{\theta\tau} &\equiv& \left[ \frac{\delta}{\delta g_{\theta\tau}} \int \sqrt{-g} d^4x g^{\mu\mu'} g^{\nu\nu'} ( R^{\lambda}_{\ \sigma\mu\nu}R^{\sigma}_{\ \lambda\mu'\nu'} ) \right]_{{\rm CA \ + \ torsionless}} \nonumber \\ 
   &=& R^{\lambda \ \theta}_{\ \sigma \ \rho}R^{\sigma\ \tau \rho}_{\ \lambda} -\frac{1}{4}g^{\theta\tau}R^{\lambda \ \xi\rho}_{\ \sigma}R^{\sigma}_{\ \lambda\xi\rho} = 0.
\end{eqnarray}

Because the solutions that we are searching are metric compatible and torsionless, then the solution $\Gamma^{\theta}_{\tau\xi}$ will become the Levi-Civita Connection for $g_{\mu\nu}$ (this is guaranteed by the Fundamental Theorem of Riemannian geometry, and note that the substitution of $\Gamma^{\theta}_{\tau\xi}$ by the Levi-Civita Connection is done only after the variation). A proper decomposition of the curvature tensor and the proper use of the Bianchi identities will convert Eq.~\ref{eq:ca_sky} and Eq.~\ref{eq:ca_step} into the Stephenson-Kilmister-Yang Equation~\cite{ref:old_2, Ref:12, Ref:13, Ref:14} and the algebraic Stephenson Equation~\cite{Ref:12,Ref:13}, respectively as,    
\begin{equation}
\label{eq:sky}
    \nabla_{\tau} R_{\xi\theta} - \nabla_{\xi}R_{\tau\theta} = 0, 
\end{equation}
\begin{eqnarray}
\label{eq:step}
   H_{\theta\tau} &=& R_{\ \sigma\theta\rho}^{\lambda} R_{\ \lambda\tau}^{\sigma \ \ \rho} -\frac{1}{4}g_{\theta\tau}R^{\lambda \ \xi\rho}_{\ \sigma}R_{\ \lambda\xi\rho}^{\sigma}  \nonumber \\
       &=& \frac{1}{2} g_{\theta\tau} R_{\sigma\rho}R^{\sigma\rho} + \frac{5}{3}R_{\theta\tau}R - 2 R_{\theta}^{\ \sigma}R_{\sigma\tau} -\frac{2}{5}g_{\theta\tau} R^2 + C_{\theta\sigma\tau}^{\ \ \ \ \rho} R_{\rho}^{\ \sigma}  \\
       &=& 0. \nonumber 
\end{eqnarray}
Here we have quoted the results given in Ref.~\cite{Ref:13}, with $C_{\theta\sigma\tau}^{\ \ \ \rho}$ as the traceless Weyl conformal curvature. 
$\nabla_{\rho}$ is the covariant derivative with respect to the Levi-Civita connections. 

Obviously the above two equations are satisfied simultaneously by the vanishing of the Ricci curvature tensor, and hence satisfied by the Schwarzschild metric. 
Note that we didn't solve Eq.~\ref{eq:var_action} directly, but routed ourselves into the domain of the theory of quadratic gravity and borrow the results from her. 
Our statement that the Schwarzschild metric is induced by a configuration of the 16 vector bosons is thus verified. 

It is also worth mentioning that there exists another metric (the Thompson-Pirani-Pavelle-Ni, TPPN, metric~\cite{ref:7, ref:9}), different from the Schwarzschild metric, that is also a simultaneous torsionless solution to Eq.~\ref{eq:sky} and Eq.~\ref{eq:step}. 
This new metric, together with the Schwarzschild metric, suggest the existence of more than one gravitational copies of matter in Nature. 
The Schwarzschild and TPPN metric are displayed in the following, respectively,
\begin{equation}
   \label{eq:S_metric}
   ds^2 = ( 1 - \frac{2GM}{r})dt^2 - (1 - \frac{2GM}{r})^{-1} dr^2 - r^2 d\Omega^2,
\end{equation}
\begin{equation}
\label{eq:TPPN_metric}
   ds^2 = ( 1 + \frac{ G'M' }{r})^{-2}dt^2 - (1 + \frac{ G'M' }{r})^{-2} dr^2 - r^2 d\Omega^2,
\end{equation}
where $GM$ and $G'M'$ are the integration constants for the solutions.

In these two solutions, the metrics happen to be compatible with their respective connections. 
But we have to bear in our mind that we have not assumed metric compatibility, as a priori, in our formulation of the theory. 
The existence of some solutions which are metric compatible does not mean that the theory is a higher derivative theory.

The TPPN solution was first discovered as a solution to the vacuum Eq.~\ref{eq:sky}, with symmetric connections and with no reference to Eq.~\ref{eq:step}. 
It was later shown by Baekler, Yasskin, Ni, and Fairchild~\cite{ref:9}, and by Hsu and Yeung~\cite{ref:9} that this solution satisfies vacuum Eq.~\ref{eq:sky} for the full connections. 
It is also easy to see that these metrics will also satisfy the vacuum Eq.~\ref{eq:step}~\cite{ref:9}. 
In fact, it was shown~\cite{ref:9} that the Schwarzschild and the TPPN solutions are the only two possible simultaneous solutions to vacuum Eq.~\ref{eq:step} and vacuum Eq.~\ref{eq:sky} for spherical symmetric situation under the compatibility ansatz. 

The most straight forward way to show the validity of the solutions given in Eq.~\ref{eq:S_metric} and Eq.~\ref{eq:TPPN_metric} is to cast vacuum Eq.~\ref{eq:step} and vacuum Eq.~\ref{eq:sky} in their spherical symmetric forms.
In the Appendix, we will give vacuum Eq.~\ref{eq:step} and vacuum Eq.~\ref{eq:sky} in their spherical symmetric forms, and we can see that both solutions satisfy the equations.

\section{The Spontaneous Breakdown of the Local Affine Symmetry to the Local Lorentz Symmetry at the Classical Gravitational Level}
A very thorny problem facing physicists when they try to develop a local affine symmetric gauge theory for gravity is the inexistence of finite dimensional spinor representations for the gauge group $GL(4\ R)$~\cite{Ref:4}. 
The observed finite dimensional spinor fields with definite spins and masses will invalidate our claim that we are having $GL(4\ R)$ symmetries in our laboratories.

This dilemma is resolved here in the following way: in spite of the fact that the Lagrangian of the theory is having the full symmetry of the affine group, the solutions of the equations of motion (for example, the classical solutions that include the Schwarzschild metric and the TPPN metric and the accelerating cosmic metric as given below) which describe our classical gravitational phenomena, retain far less symmetries.
      
What are the residual symmetries that are left unbroken? 
This can be answered by noting that the classical $\Gamma^{\rho}_{\tau\mu}$ are now the Levi-Civita Connections, which are compatible with the background metric. 
This compatibility will require that the observed Yang-Mills vector fields $A^{m}_{\ n\mu}$ be anti-symmetric in their $m$, $n$ indices. 
For this case, the ten symmetric generators $T^{ab}$ of the $GL(4\ R)$ will not be used. 
The remaining six anti-symmetric generators $J^{ab}$ are just the generators of the local Lorentz Group. 
The geometric picture for the Compatibility Ansatz is thus that we are going to identify the defining Minkowskian frame for the vierbein fields and its rotations as our admissible geometric settings in the discussions of classical gravity physics.  

Hence what is left unbroken by our solutions is the local Lorentz symmetry. 
The changes in the gauge potentials are compensated by the changes in the vierbein fields when the Minkowskian axes are rotated, so as to leave our metrics and Levi-Civita Connections unchanged. 
That explains why we are now seeing particles of definite spins and masses in our laboratories. 
In summary, classical gravity is expressed by a spontaneously broken Erlangen program~\cite{Ref:18}, breaking down the local $GL(4\ R)$ symmetry to the local Lorentz symmetry. 

Some may wonder if it is legitimate to use $GL(4\ R)$ for a local gauge theory became of its non-compactness. 
The reader is referred to the Appendix of Ref.~\cite{ref:10} for a discussion on the positive-definiteness of $GL(4\ R)$.

\section{Applications on Dark Matter and Dark Energy problems}
\subsection{An interpretation of the two legitimate solutions}
Though the Schwarzschild metric has already found a lot of applications in the studies of various gravitational phenomena, the TPPN metric was dismissed by its discoverers soon after its discovery because it failed to reproduce the classical tests that are so successfully predicted by the Schwarzschild metric. 
And this leads, subsequently by many people, to the conclusion that the Yang-Mills gauge theory is not a viable physical theory.

Here we want to show that a suitable interpretation of the above two metrics can save this situation. And will reproduce nicely both the galactic rotation curves and the amounts of intergalactic lensing, if Nature follows our following postulate.

We postulate that if Nature is going to make use of this Yang-Mills gauge theory of gravity, matter will then be endowed with either one of these two metrics. 
Matter endowed with the first metric ($\bar{g}$, metric given in Eq.~\ref{eq:S_metric}) will be called the regular matter, and matter endowed with the second metric ($g'$, metric given in Eq.~\ref{eq:TPPN_metric}) will be called the primed matter. 
Both the regular matter and the primed matter were produced during the creation of the Universe, though in different amounts, probably because of the difference in the requirements of energy in producing them. 

There might possibly be other particles endowed with some yet undiscovered metric solutions. 
These particles are assumed to be too heavy to be produced in an amount large enough to be noticeable in the present day astronomical observations. 
In other words, we will suffice ourselves by considering the physics played by these two gravitational copies of matter. 

The immediate question is: how do we know that there exist solution to Eq.~\ref{eq:S_metric} and Eq.~\ref{eq:TPPN_metric} that describe the co-existence of the regular matter $M$ and primed matter $M'$ as the source of the gravitational field? 

The gravitational field equations are highly nonlinear, a rigorous solution for a source with both regular matter $M$ and primed matter $M'$ is hard to find, but we can infer the existence of such a solution by forming a linear superposition of 
\begin{equation}
  \label{eq:superposition}
  g = \frac{GM}{GM + G'M'} \bar{g} + \frac{G'M'}{GM + GM'} g'. 
\end{equation}
This linearly superposed metric is asymptotically Minkowskian. 
As we are going to explain in the Appendix, the fact that $GM$ and $GM'$ are integration constants for the solutions will imply that they do not appear in the equations that embody the solutions. 
In the limit of $\frac{GM}{G'M'} \to 0$ (or $\frac{G'M'}{GM} \to 0$), $g$ will be getting closer and closer to be a solution. 
The deviations of $g$ from a solution will be proportional to $\frac{GM}{G'M'}$ or its higher orders as $\frac{GM}{G'M'} \to 0$ (or to $\frac{G'M'}{GM}$ or its higher orders as $\frac{G'M'}{GM} \to 0$). 
Note that $\frac{GM}{GM + G'M'}$ signifies the significance of the Schwarzshild component in $g$ while $\frac{G'M'}{GM+G'M'}$ signifies the significance of the TPPN metric in $g$. More details on this can be found in the Appendix. 

The next question is: how does a test particle, which is either a regular matter of mass $m$ or a primed matter of mass $m'$ reacts to the gravitational field which is produced by a mixture of $M$ and $M'$?

The energy-momentum tensor $T_{\theta\tau}(g)$ is a functional of the metric. 
It is natural to assume that when it is evaluated at $\bar{g}$, it will be the energy-momentum tensor for the regular matter. 
And when it is evaluated at $g'$, it will be the energy-momentum tensor for the primed matter. 
So we have
\begin{eqnarray}
\label{eq:eng_mom_tensors}
    {\rm energy{\textrm -}momentum\ tensor\ for\ the\ regular\ matter} = T_{\theta\tau}(\bar{g}),       \nonumber     \\ 
    {\rm energy{\textrm -}momentum\ tensor\ for\ the\ primed\ matter} = T_{\theta\tau}(g').  
\end{eqnarray}

It was explicitly pointed out by Stephenson~\cite{Ref:12} that when we use the full connections (these connections are having both the symmetric and the anti-symmetric parts) as dynamical variables, then the Bianchi identities and vacuum Eq.~\ref{eq:sky} together will imply a local covariant conservation law for the metric energy-momentum tensor  
\begin{equation}
\label{eq:dev_eng_mom_tensor}
   \nabla^{\theta}(\Gamma)H_{\theta\tau} = \nabla^{\theta}(\Gamma)T_{\theta\tau}(g) = 0.
\end{equation}
This is valid for any generic metric and connections which satisfy vacuum Eq.~\ref{eq:sky}. 
In particular when the metric is $\bar{g}$ (or $g'$) with corresponding connection $\bar{\Gamma}$ (or $\Gamma'$), we will have two separate conservations laws
\begin{eqnarray}
\label{eq:sep_conser}
    \nabla^{\theta}(\bar{\Gamma})T_{\theta\tau}(\bar{g}) = 0,  \nonumber \\
    \nabla^{\theta}(\Gamma')T_{\theta\tau}(g') = 0.
\end{eqnarray}
Eq.~\ref{eq:eng_mom_tensors} and Eq.~\ref{eq:sep_conser} together will mean that the regular matter and the primed matter are separately conserved.
Therefore, a regular (primed) particle at $x$ will mean an energy-momentum tensor that is localized at $x$ and satisfies the first (second) conservation law. 
     
By integrating over a small spacetime region surrounding the test particle, a local covariant conservation law can be translated into an equation of motion for the test object which shares the same $T_{\theta\tau}$ with the source but not affecting the metric~\cite{ref:11}. 
Eq.~\ref{eq:eng_mom_tensors} and Eq.~\ref{eq:sep_conser} will mean that the regular (primed) matter will move under the influence of the regular (primed) metric which is generated by regular (primed) matter.
In other words the regular (primed) matter will interact only with the regular (primed) matter gravitationally.
 
A little care should be paid in the situation of a source of a mixture of $M$ and $M'$: 
In the presence of a metric given in Eq.~\ref{eq:superposition}, a regular test particle will have an energy-momentum tensor of the form of $T_{\theta\tau}(\frac{GM}{GM+G'M'}\bar{g})$. 
Because either $\bar{g}$ or $\frac{GM}{GM+G'M'}\bar{g}$ gives the same $\bar{\Gamma}$ ($\bar{\Gamma}$ won't change under a constant scaling of the metric), $T_{\theta\tau}(\frac{GM}{GM+G'M'}\bar{g})$ will obey the same local covariant conservation law as $T_{\theta\tau}(\bar{g})$. 
Similar situation holds for a primed test particle. 

Other than gravitational interactions, the regular matter and the primed matter are assumed to be identically the same in all other interactions.

The geodesic equations are equivalent to the acceleration equations under the respective metric. 
And the respective acceleration produced by these metrics, when the speed of motion is small compared with the speed of light, can be calculated from
\begin{equation}
\label{eq:accel}
  \frac{d^2 r}{dt^2} = \frac{1}{2}g^{rr}\frac{\partial g_{00}}{\partial r},
\end{equation}
and are respectively $-\frac{GM}{r^2}$ and $-\frac{G'M'}{r^2}( 1 + \frac{G'M'}{r})^{-1}$.

The $G'$ and $M'$ are introduced in order to make a parallel comparison between the Newtonian gravitational force and the new gravitational force. 
We will call $G'$ the primed gravitational constant, and $M'$ the primed gravitational mass. 
This new Gravitational force is attractive when $G'M'$ is positive.

Let us now consider the interesting situation in which a regular matter of mass $m$ is sticking together with a primed matter of mass $m'$ through non-gravitational interactions, and in which they are moving together in a circular motion of radius $r$ under the influence of a gravitation field produced by a source consisting of a regular mass $M$ and a primed $M'$. 

Note that even though the regular matter and the primed matter may be very close together, they nevertheless occupy different spacetime points and hence their energy-momenta will not overlap, and hence they will interact with their own gravitational forces, and the centrifugal force will balance the combined gravitational forces 
\begin{equation}
\label{eq:centri_force}
  (m+m')\frac{v^2}{r} = \frac{GM}{r^2}m + \frac{G'M'}{r^2}( 1 + \frac{G'M'}{r})^{-1}m', 
\end{equation}
and then we will arrive at a relationship holding between the rotation speed $v$ with the distance $r$,
\begin{equation}
  \label{eq:v_squ}  
  v^2 = \frac{GM}{r}\frac{m}{m+m'} + \frac{G'M'}{r+G'M'}\frac{m'}{m+m'}.
\end{equation}

\subsection{Predictions of the Universal rotation curves for spiral galaxies}     
There is an immediate application of Eq.~\ref{eq:v_squ} to describe the kinematics of a stellar object which is made up of the regular matter $m$ and the primed matter $m'$, and is moving at a distance $r$ from the center of a spiral galaxy. 
The spiral galaxy is assumed to be composed of a disk solely of regular matter of mass $M$ and a halo composed of a uniformly distributed matter with a contamination of primed matter of mass density $\rho'$. Because of 
\begin{equation}
  \label{eq:M_primed}  
  M' = \frac{4\pi}{3}r^3 \rho',
\end{equation}
where $M'$ is the total primed mass inside $r$, and because the arm structure of the regular matter presents its contribution in modified Bessel functions form~\cite{ref:12}, the total contributions to $v^{2}$ by the spiral arms and the halo will sum up to
\begin{eqnarray}
    \label{eq:v_squ_new}
v^2 &=& v^2_d(\frac{m}{m+m'}) + \frac{ G'\frac{4\pi}{3}\rho'r^3 }{ r + G'\frac{4\pi}{3}\rho'r^3 }(\frac{m'}{m+m'}) \nonumber\\
    &=& 4\pi G \Sigma_0 h y^2 [I_0(y)K_0(y) - I_1(y)K_1(y) ] \times (\frac{m}{m+m'}) \\ 
    && + \frac{ G^{\star} r^3}{r + G^{\star}r^3 }(\frac{m'}{m+m'}), \nonumber
\end{eqnarray}
where $I_{i}$, $K_{i}$ are modified Bessel functions of the first and second kinds.
The $v_d^2$ is the speed coming from the Newtonian force of the disk with $y= \frac{r}{2h}$~\cite{ref:12}.

\begin{figure*}[htbp]
  \begin{center}
    \subfigure[]{
      \label{fig:RC_MilkyWay}
      \includegraphics[width=0.47\textwidth]{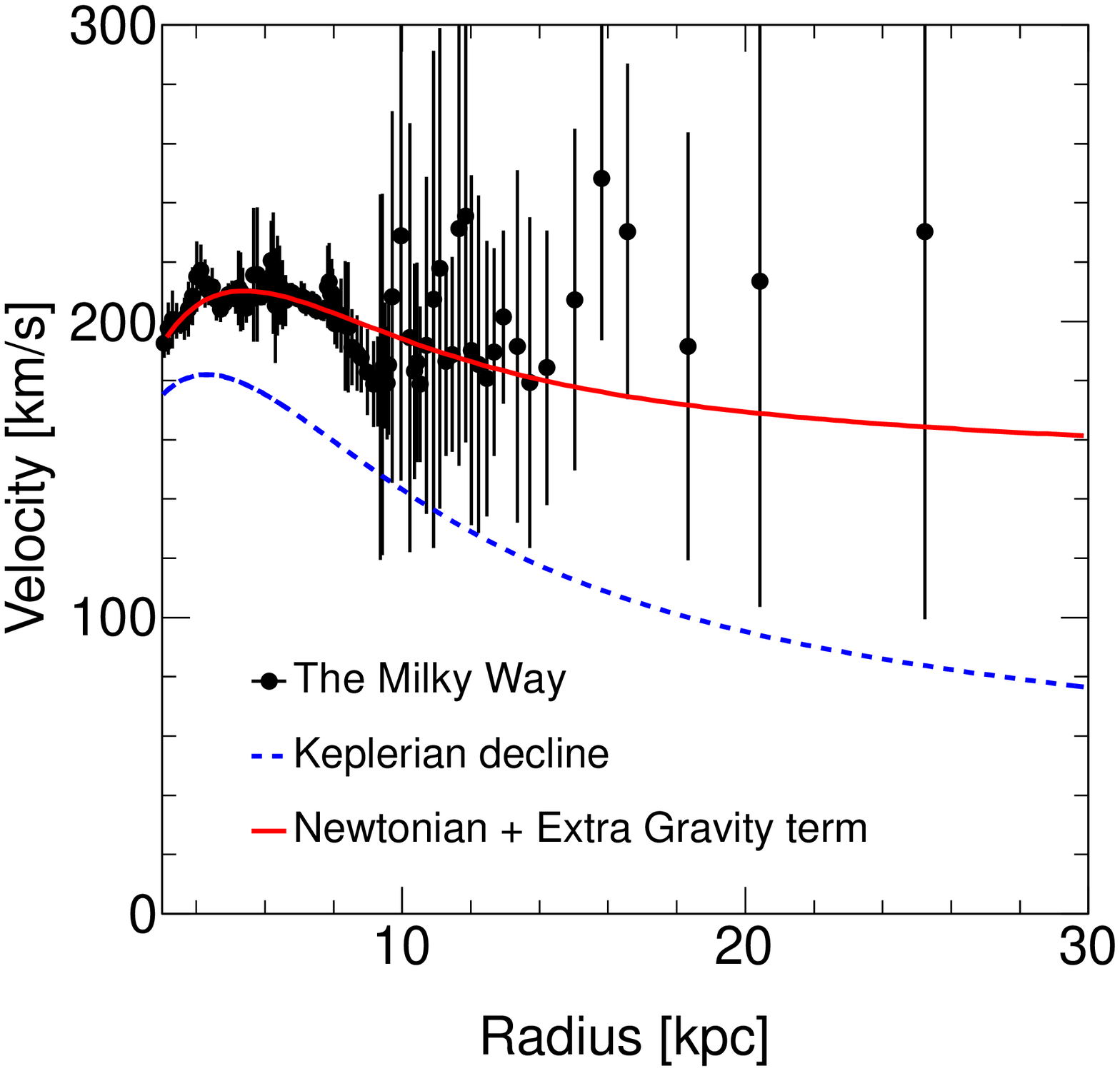}
    }
    \subfigure[]{
      \label{fig:RC_3198}
      \includegraphics[width=0.47\textwidth]{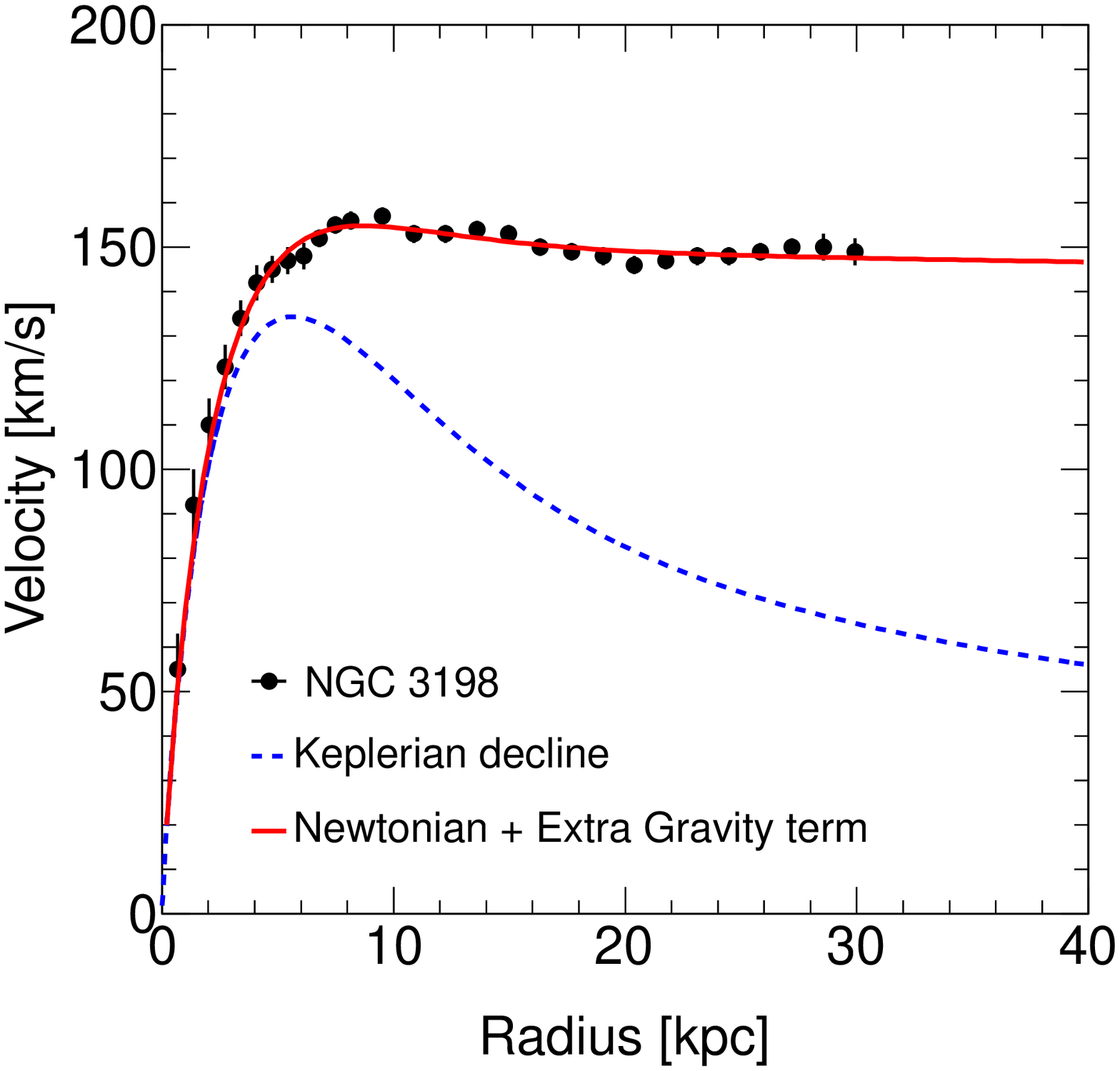}
    }
    \subfigure[]{
      \label{fig:RC_2403}
      \includegraphics[width=0.47\textwidth]{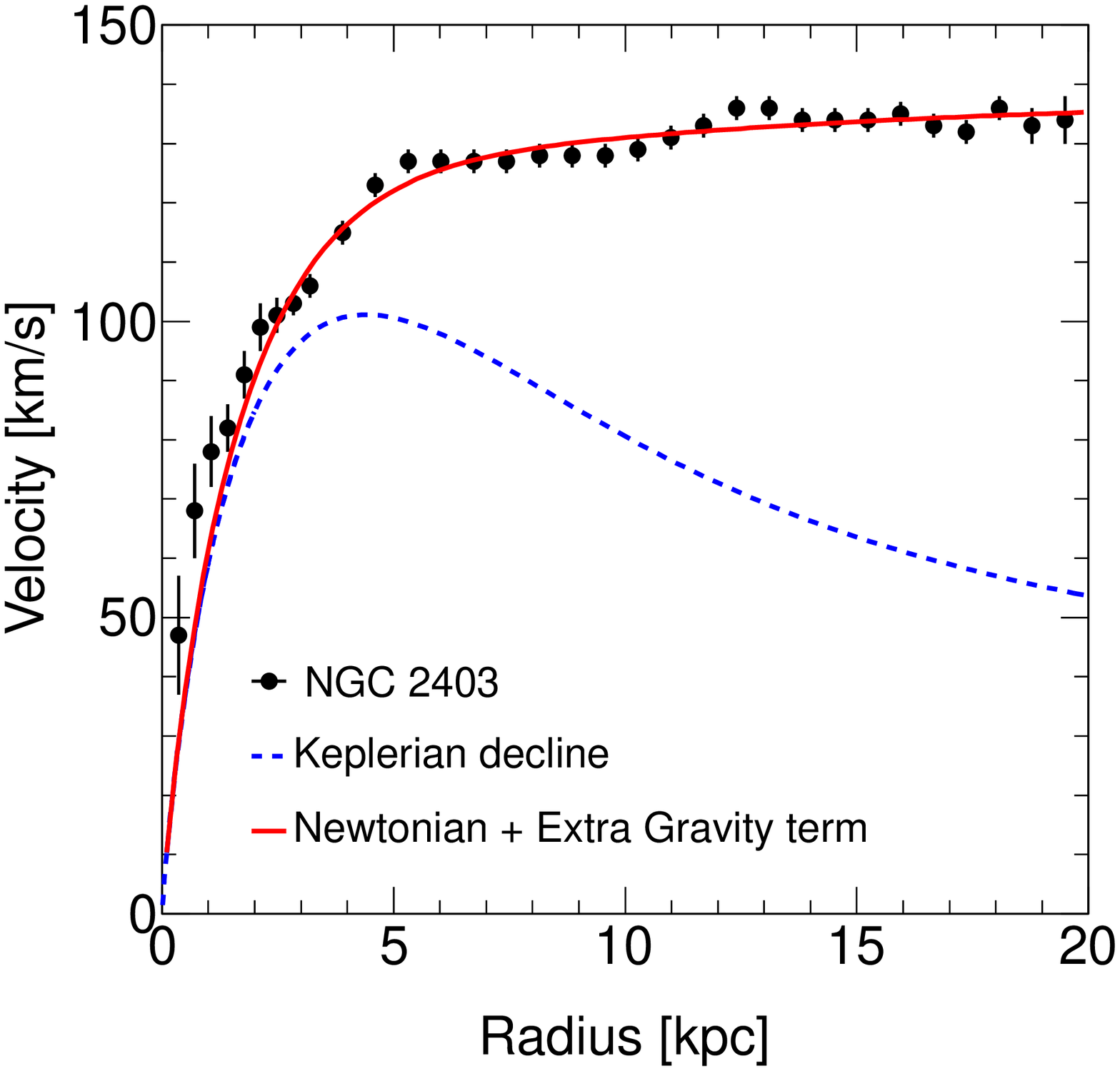}
    }
    \subfigure[]{
      \label{fig:RC_6503}
      \includegraphics[width=0.47\textwidth]{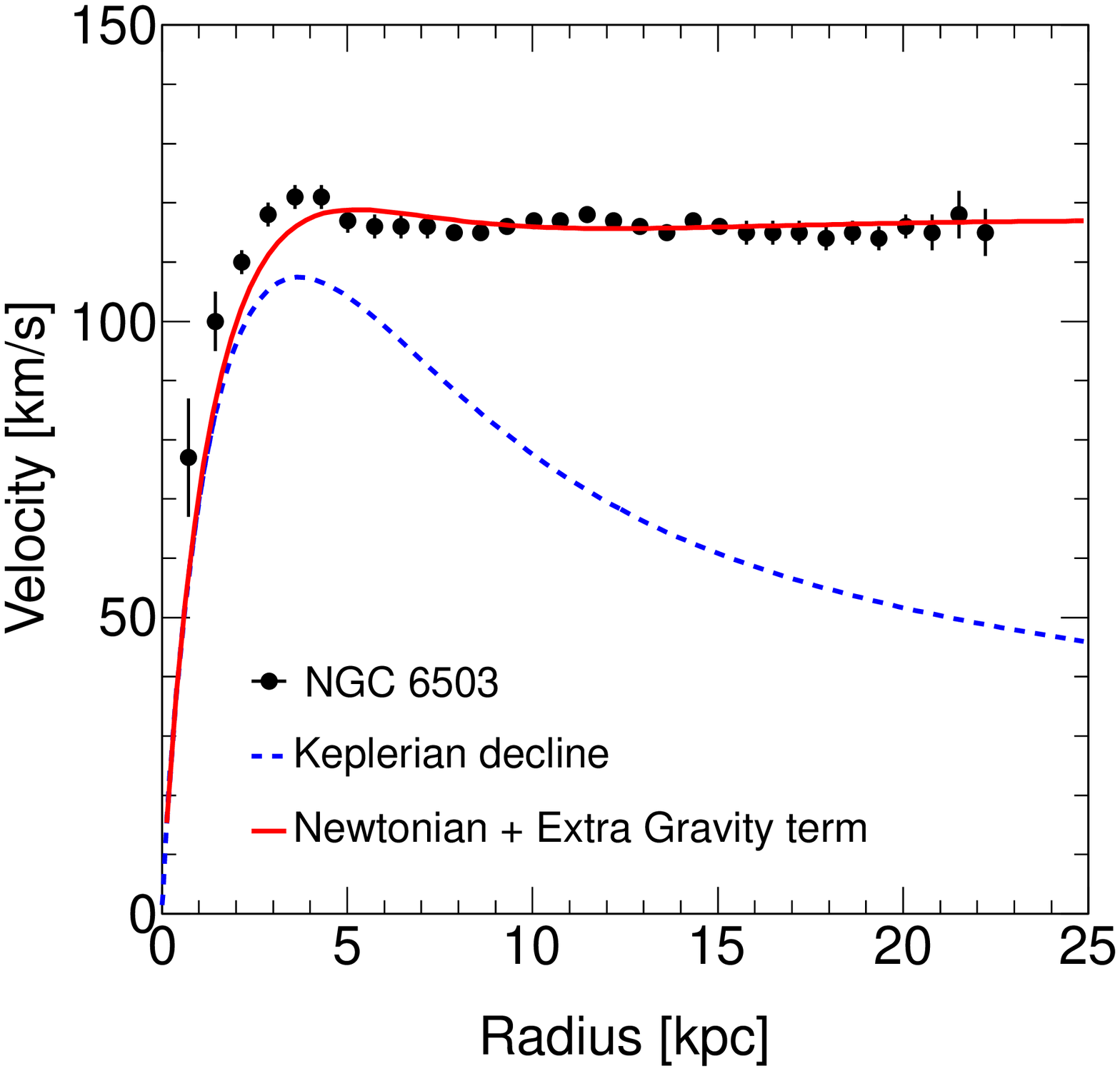}
    }
    \end{center}
  \caption{  The predicted relationship between the galactic rotation speed $v$ and the distance $r$ from a combined influence of the Newtonian force and the new gravitational force: \subref{fig:RC_MilkyWay}The Milky Way, \subref{fig:RC_3198}NGC 3198, \subref{fig:RC_2403}NGC 2403 and \subref{fig:RC_6503}NGC 6503. 
  \label{fig:1}}
\end{figure*}

We will see immediately that this agrees exactly with the empirical formula given by Salucci $et$ $al$~\cite{ref:13} who analyzed a large number of spiral galaxies and drew up a formula to describe their rotation curves. 
We have also extracted the values of $G^{\star}$ (which is the product of the primed gravitational constant and the primed matter density in the halo) and $m'/m$ (which is the ratio of the primed mass density to the regular mass density in the halo) by fitting some well-known galaxies~\cite{ref:14,ref:15,ref:16,ref:17,ref:18}, and it is amazing to find that a universal value of $G^{\star} \approx 10^{-2}\ \mathrm{kpc^{-2}}$ and a universal ratio of $\frac{m'}{m} \approx \ 2 \times10^{-9}$ fit very well with the observed results when $r$ ranges from 3 kpc to 30 kpc. Note also that the value of $G\Sigma_0$ and $h$ are more or less the same as those observed.
The results are shown in Fig.~\ref{fig:1} and Table~\ref{table:1}. 

\begin{figure}[htbp]
  \begin{center}
      \includegraphics[width=0.47\textwidth]{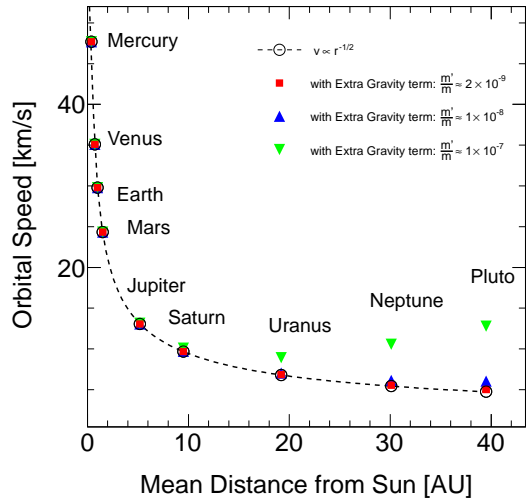}
  \end{center}
  \caption{ The orbital speeds of the planets in the solar system predicted by Eq.~\ref{eq:v_squ_new} with different $m'/m$ values. $2\times10^{-9}$, $1\times10^{-8}$ and $1\times10^{-7}$ are for red, blue and green points respectively. 
      \label{fig:2}}
\end{figure}

\begin{table*}[htbp]
   \begin{center}
   \begin{tabular}{ccccc}
    \hline
    \hline
                                        &    The Milky Way    &     NGC 3198       &     NGC 2403       &     NGC 6503        \\
    \hline
     $G\Sigma_0$ [$km^2s^{-2}kpc^{-1}$] &  $6.8\times10^{3}$  & $2.8\times10^{3}$  & $2.1\times10^{3}$  &  $2.8\times10^{3}$   \\
      h  [kpc]                          &        2.0          &   2.63             &      2.05          &      1.72            \\
     $G^{\star}$  [kpc$^{-2}$]          &  $5.0\times10^{-2}$ & $9.2\times10^{-3}$ & $1.4\times10^{-2}$ &  $1.3\times10^{-2}$  \\
     $\frac{m'}{m}$                     &  $2.3\times10^{-9}$ & $2.2\times10^{-9}$ & $2.0\times10^{-9}$ &  $1.4\times10^{-9}$  \\
    \hline
     \hline
   \end{tabular}
    \end{center}
   \caption{The fitting parameters for some galaxies.}
\label{table:1}
\end{table*}
We also have to check for the influence of this diffuse halo medium on the motion of the planets in our solar system. 
With the values of the $G^{\star}$ and $m'/m$ given in the above, the observed rotation speeds of the planets fit well with the predictions of Eq.~\ref{eq:v_squ_new}. 
Figure~\ref{fig:2} gives the planetary motions for various values of $G^{\star}$ and $m'/m$.
     
The alert reader may find that, in the above discussions, we have already made the assumption that the stellar objects in the galaxies are composed solely of regular matter.
This assumption can be understood in the following way. 
The primed matter always respond to the primed gravitational pull with a high rotation speed even when they are bound with some regular matter and hence are far harder for them to condense gravitationally into a star.

Though the stars contain no primed matter, the stars that are rotating at the outskirt of a spiral galaxy are, in fact, embedded in pockets of halo media and are rotating around the galaxy all together.  
     
The amount of regular matter in the halo is small when compared with that in the galactic bulge and spiral arms, and is thus neglected in the above discussions.

\subsection{Right amounts of intergalactic gravitational lensing}
Next, let us turn to see what the primed matter does in explaining the large light deflections that are observed in intergalactic gravitational lensing. 
We shall take the galaxy cluster Abell 1689 as our illustration. 
We shall regard Abell 1689 as a cluster consisting of galaxies which are carrying their own individual halos with them. 
And this collection of galactic halos forms the halo of the cluster. 
Since the galactic halos are always regarded as having a size of the order of 30 kpc, the size of the cluster will be very close to its halo size which is taken to be 300 kpc.

The azimuthal angle swept by the light, when it travels from point $R$ to the point of closest approach $r_0$, under the influence of gravity described by the metric
\begin{equation}
  \label{eq:gen_metric}  
  ds^2 = B(r)dt^2 - A(r)dr^2 -r^2d\Omega^2,
\end{equation}
is given by~\cite{ref:19}
\begin{equation}
  \label{eq:delta_phi}  
  \Delta\varphi \equiv \varphi(r_0) - \varphi(R) = \int^{R}_{r_0} A^{\frac{1}{2}}(r)[(\frac{r}{r_0})^2 \frac{B(r_0)}{B(r)} -1 ]^{-\frac{1}{2}} \frac{dr}{r}.
\end{equation}

In the case of the TPPN metric for a point source of primed mass $M'$, the angle swept is
\begin{equation}
  \label{eq:delta_phi_integral}  
  \Delta\varphi = \int^{R}_{r_0} \frac{(r_0 + G'M')dr}{(r+G'M')[(r+G'M')^2 - (r_0+G'M')^2]^{\frac{1}{2}}},
\end{equation}
which can be readily integrated to give
\begin{equation}
  \label{eq:delta_phi_result}  
  \Delta\varphi = \sec^{-1} \frac{1+\beta R}{1+\beta r_0},
\end{equation}
with $\beta$ = $\frac{1}{G'M'}$.

We will immediately notice that the angle change will be $\frac{\pi}{2}$ when $R$ goes to infinity. 
That means no deflection for the light by a point source of the primed matter when it comes from infinity and then goes back to infinity. 
Things will be different when we deal with a distributed source of the primed matter as we are going to show in the following.

Let $R$ be the radius of the halo of Abell 1689 and $r_0$ be the closest approach from the cluster center. 
The light will see a point source of the primed matter of constant mass $M'$ when it is moving beyond the halo. 
And it will see a point source of diminishing mass (i.e. increasing $\beta$) when it enters the halo because it sees only the mass that lies inside $r$.

We claim that the deflection by a point source of mass $M'$ is smaller than that by a uniformly distributed halo which has a total mass of $M'$, if the light penetrates into the halo during some time in its journey. The above claim is obvious by noting that 
\begin{equation}
  \label{eq:delta_phi_result_1}  
  \frac{d(\Delta\varphi)}{d\beta} > 0.
\end{equation}
So if the light penetrates into the cluster halo, the total azimuthal angle change will be greater than $\frac{\pi}{2}$, and we will see light deflectng towards the center of the cluster.

The actual angle swept, when light comes from the infinity, enters the halo at $R$ and reaches the point of closest approach at $r_0$ is given by 
\begin{equation}
  \label{eq:delta_phi_result_2}    
  \frac{\pi}{2} + ( \Delta\varphi|_{\rm{halo}} - \Delta\varphi|_{\rm{point\ source}} ).
\end{equation}
The second term in the bracket in Eq.~\ref{eq:delta_phi_result_2} comes from the fact that we have to take out a $\frac{\pi}{2}$ in calculating the angle of deflection.
The quantity inside the bracket of Eq.~\ref{eq:delta_phi_result_2} can be regarded as a change $\delta(\Delta\varphi)$ in $\Delta\varphi$ due to a change $\delta M'$ in $M'$, and we get  
\begin{eqnarray}
    \label{eq:delta_phi_result_3}
 &&\Delta\varphi|_{\rm{halo}} - \Delta\varphi|_{\rm{point\ source}} \nonumber \\ 
  && = \delta(\Delta\varphi) \nonumber\\
  && = \frac{\partial (\Delta\varphi)}{\partial M'}\delta M'   \\
  && = \frac{\partial}{\partial M'} [\sec^{-1} \frac{1+\beta R}{1+\beta r_0}] \delta M' \nonumber \\ 
  && = \frac{3}{\sqrt{2}} (R - r_0)^{ \frac{3}{2}} G^{\star -\frac{1}{2}} R^{-\frac{5}{2}}.  \nonumber 
\end{eqnarray}

A plug in the data of $R$ = 300 kpc, $(R - r_0)$ = 30 kpc and $G^{\star}$ = $10^{-2}$ $\mathrm{kpc^{-2}}$, will give us an angle of deflection, which is 2$\delta (\Delta \varphi)$, a value of $4\times 10^{-3}$, which is many times of that expected from General Relativity.
 
Note that the value of $G^{\star}$ that we used in calculating the Abell 1689 light deflection comes from the curve fittings of galactic rotations. 
Again there seems to be a universal value for $G^{\star}$, as it should be. 

We should also note that the sun and the planets in the solar system aren't contaminated with the primed matter as we have explained in the above. 
The fundamental tests on General Relativity will hence remain intact.

\subsection{Primordial torsion and the accelerating expansion of the Universe}
There is another nice feature of this Yang-Mills gauge theory of gravity, when we use it to study the Universe as a whole. 
This theory was shown to admit a cosmological solution of the form~\cite{ref:20, ref:21}
\begin{equation}
\label{eq:metic_cosmology}
 ds^2 = -dt^2 + r_0^2 e^{2 \xi t} ( dr^2 + r^2d\theta^2 + r^2\sin^2\theta d\phi^2), 
\end{equation}
with primordial local torsion compoments, 
\begin{equation}
\label{eq:non_zero_torsions}
  \tau_{\hat{0}\hat{1}\hat{1}} = \tau_{\hat{0}\hat{2}\hat{2}} = \tau_{\hat{0}\hat{3}\hat{3}} = \frac{\xi}{2}. 
\end{equation}

The $r_0$ and $\xi$, with $ \xi > 0$ are integration constants arising from the integration of the equation of motion. 
We can interpret this solution as representing an expanding and accelerating Universe when the influence of gravity dominates over the influence of matter and radiation.
The role played by the primordial torsion is crucial here: the stretching on the Universe by the metric is compensated by the twisting by the torsion. 
And from the metrical point of view, we look like living in a Universe with a cosmological constant $\xi$.  
And interesting enough, our torsion selects the spatially flat metric ($\kappa$ = 0) as the only accompanying metric~\cite{ref:20, ref:21}.
The spatially flat geometry of the Universe is confirmed by WMAP. 
For more information on the predictions of the Yang-Mills type gauge theory of gravity on the evolution of the primordial Universe, the reader is referred to Ref.~\cite{ref:22}.


\section{Discussions}
We are fully aware that some of the terms and ideas used here have already appeared in the literatures. 
But we want to emphasize that they are appearing here in very different contexts. 
For example, we do not regard the world metric in our theory as a fundamental variable. 
Instead, the world metric is just an arbitrary background of measuring clock and stick for our spacetime. 
The observed metric takes a particular form (for example the Schwarzschild metric) simply because the affine gauge bosons require that particular metric in order to exist as a solution. 
As a result, we will not have spin-2 gravitons in our theory; the 16 spin-1 vector bosons are the only propagating particles for gravitation. 
And the dynamics of gravity is fully governed by the $GL(4\ R)$ Yang-Mills equation of motion. 
Our theory is therefore a theory of vector gravity. 

It might be interesting to point out that there had been many works done in the direction called the gauge theories of gravity. 
Yet most of them are not gauge theories in the sense of Yang and Mills. 
People are either reluctant to give up the metric (or the vierbein) as gauge potential because they might think that the metric is too important for gravity theory to be put in an auxiliary position. 
Or they might be afraid to take up the gravitational Lagrangian that is quadratic in the field strength tensor as Yang and Mills did, because of the fear that there could be spurious solutions that will upset the known gravitational observations~\cite{Ref:16}. 
But an important factor that is preventing people to arrive earlier at a true gauge theory of gravitation based on $GL(4\ R)$, as we believe, is the wrong perception that the particle contents of the theory will not fit into experiments. 
It is one of our observations, as we have explained in the above, that our observed gravitational world is, in fact, a classical solution to the affine symmetric theory and is in a state of spontaneously broken symmetry. 
The remaining symmetry is the local Lorentz symmetry. 

Finally, a gravity theory with the Lorentz group (which is a subgroup of $GL(4\ R)$) as the local gauge group can be obtained from our theory by simply restricting ourselves to the anti-symmetric components of the $A^{m}_{\ n\mu}$. 
Only the generators $J^{ab}$ and the first commutation relation of Eq.~\ref{eq:commu_rel} will be used. 
And we shall have a Yang-Mills gauge theory of 6 gauge vector bosons. 
All the results given in this article will then remain valid, with the exceptions that the local Lorentz symmetry will now be honored by both the Lagrangian and the solutions to the equations of motion, and that the metric and the connections will now be compatible automatically. 
One price has to be paid, though, by gauging the Lorentz group instead of the $GL(4\ R)$. 
The Riemannian tensor will then contain terms which are of second derivatives in the metric, and the metric and the torsion (which are functions of the vierbeins and the gauge potentials) are independent variables. 
And gravity theories basing on Lagrangians quadratic in the Riemannian tensor may have to face the affections by many of the pathologies of higher derivative theories. 
From the solutions of the equation of motion, it seems that gauging the Lorentz group is good enough to describe the present day physical phenomena of gravity. 
Apart from staying away from the pathologies that we have just mentioned by gauging the Lorentz group, we are venturing into $GL(4\ R)$ because we believe that there might be something in physics, in addition to gravity, that can be explained by the full affine symmetric gauge theory~\cite{Ref:4,Ref:19}.    

The Yang-Mills gauge theory of gravity has a richer structure than that of Einstein’s General Theory of Relativity. 
Its richer number of solutions, both in the absence and in the presence of torsion allow us to describe more physical phenomena with it. 
The recently observed astronomical Dark Matter and Dark Energy phenomena seems to show that Nature is enjoying the full use of this Yang-Mills gauge theory of gravity.

\appendix
\setcounter{equation}{0}
\section*{Appendix The Static Spherically Symmetric Gravitational Equations and Their Solutions}
\refstepcounter{section}
In this appendix we will cast the vaccum Stephenson Equation and the vacuum Stephenson-Kilmister-Yang Equation in static spherical symmetric forms.
The metric is taken to be the form of Eq.~\ref{eq:gen_metric}, namely 
\begin{equation}
  \label{eq:gen_metric_2}  
  ds^2 = B(r)dt^2 - A(r)dr^2 -r^2d\Omega^2.
\end{equation}
There are three independent, non-vanishing components for the left-hand-side of Eq.~\ref{eq:step}. 
The $tt$ component, 
\begin{eqnarray}
    \label{eq:tt_component}
   && 8r^2A^2B^2B'^2 + r^4(A'B'B - AB'^2 - 2AB'' + 2AB'^2)^2 \nonumber  \\
   &&     - 16A^2B^4(A-1)^2 - 8r^2A'^2B^4 = 0, 
\end{eqnarray}
the $rr$ component,
\begin{eqnarray}
    \label{eq:rr_component}
   && 8r^2A'^2B^4 + r^4(A'B'B - AB'^2 - 2ABB'' + 2AB'^2)^2  \nonumber  \\
   && - 16A^2B^4(A-1)^2 - 8r^2A^2B^2B'^2 = 0,
\end{eqnarray}
the $\theta\theta$ component (same as the $\phi\phi$ component),
\begin{eqnarray}
    \label{eq:thetatheta_component}
    && r^4(A'B'B - AB'^2 - 2ABB'' + 2AB'^2)^2  \nonumber  \\
    && - 16A^2B^4(A-1)^2 = 0.
\end{eqnarray}
There are two independent components for the left-hand-side of Eq.~\ref{eq:sky}. 
They are~\cite{ref:7}
\begin{eqnarray}
    \label{eq:SKY_ind_1}
    &&r^2(2A^2B^2B^{\prime\prime\prime} - 4A^2B''B'B - 3 AA'B^2B'' + 2A^2B'^3  \nonumber  \\ 
    &&    + 2AA'BB'^2 - AA''B^2B' + 2A'^2B^2B' )  \nonumber  \\
    &&    + 2rAB(2ABB'' - AB'^2 - A'BB') - 4A^2B^2B'  \nonumber  \\
    && = 0,
\end{eqnarray}
and 
\begin{eqnarray}
    \label{eq:SKY_ind_2}
    r^2(2AA''B^2 + AA'BB' - 4A'^2B^2) - 4A^2B^2(A-1) =  0.
\end{eqnarray}
Note that all these equations contain terms that are polynomials in $A$ and $B$. 

Subtracting Eq.~\ref{eq:rr_component} from Eq.~\ref{eq:tt_component} will give 
\begin{eqnarray}
    \label{eq:subtract_eq}
    8r^2A^2B^2B'^2 - 8r^2A'^{2}B^4 = 8r^2A'^2B^4 - 8r^2A^2B^2B'^2
\end{eqnarray}
and will lead to 
\begin{eqnarray}
    \label{eq:subtract_eq_2}
    \frac{B'}{B} = \pm \frac{A'}{A}.
\end{eqnarray}

The primes appearing as suffices in Eq.~\ref{eq:gen_metric_2} to Eq.~\ref{eq:subtract_eq_2} signify differentiations with respect to $r$.

Eq.~\ref{eq:subtract_eq_2} are the two solutions for the Schwarzchild metric and the TPPN metric, respectively. 

For halos with both regular matter and primed matter whose densities are decreasing as $r$ increases, the RHS of Eq.~\ref{eq:gen_metric_2} to Eq.~\ref{eq:SKY_ind_2} will no longer be vanishing, and Eq.~\ref{eq:subtract_eq_2} will not hold. 
In that case, as explained in the following, an approximation to the solutions of the Gravitational Equations can be constructed as the superposition of $\bar{g}$ and $g'$ in the form of
\begin{eqnarray}
    \label{eq:superposition_app}
    g = \frac{GM}{GM+G'M'}\bar{g} + \frac{GM'}{GM+G'M'}g',
\end{eqnarray}
and if we use $B(r)$ and $A(r)$ to denote the $tt$ and $rr$ components of the metric of Eq.~\ref{eq:superposition_app} then
\begin{eqnarray}
    \label{eq:B_and_D}
    B(r) = \bar{B} + \frac{G'M'}{GM}B' \nonumber \\
    A(r) = \bar{A} + \frac{G'M'}{GM}A'
\end{eqnarray}
in the limit of $\frac{GM'}{GM} \to 0$. 
The bars and primes here are used to label the regular matter and the primed matter as we have mentioned in the article. 
Note that the above equations of Eq.~\ref{eq:gen_metric_2} to \ref{eq:SKY_ind_2} do not have $GM$ and $GM'$, which appear in Eq.~\ref{eq:superposition_app}, as their coefficients because $GM$ and $GM'$ are integration constants of the solutions. 
Substituting Eq.~\ref{eq:B_and_D} into Eq.~\ref{eq:tt_component},~\ref{eq:rr_component},~\ref{eq:thetatheta_component},~\ref{eq:SKY_ind_1} and~\ref{eq:SKY_ind_2}, and if we remember that $\bar{B}$ and $\bar{A}$ satisfy the original equations, then the left-hand-side of Eq.~\ref{eq:tt_component},~\ref{eq:rr_component},~\ref{eq:thetatheta_component},~\ref{eq:SKY_ind_1} and~\ref{eq:SKY_ind_2} will contain only terms that are proportional to $\frac{G'M'}{GM}$ or it higher orders. 
What that means is that for some given $GM$ and $G'M'$, and some given $\bar{B}$ and $\bar{A}$ satisfying the original equations, $B(r)$ and $A(r)$ will satisfy the equations  approximately as far as $\frac{G'M'}{GM}$ is very small. 
That means the superposition given in Eq.~\ref{eq:superposition_app} is an approximate solution to the equations as far as $\frac{G'M'}{GM}$ is very small. 
The physical consequence is: at least in the limit of $\frac{G'M'}{GM}\to 0$, a solution of the form of Eq.~\ref{eq:superposition_app} which represents halos having mixtures of regular and primed matter exist in nature. 
Similar arguments apply to the case when $\frac{GM}{G'M'}$ is very small.

\section*{Acknowledgments}
We would like to thank Professor Friedrich W. Hehl, Professor James M. Nester, Professor T. C. Yuan, and Professor Daniel Wilkins for various suggestions. 
In particular, we would like to thank Professor Nester for pointing out some errors in the previous version of our manuscript.



\begin{thebibliography}{10}
\bibitem{Ref:9} Eq. 16 looks like the relations between the spin connections and the Levi-Civita Connections under the tetrad postulate. Here we want to emphasize that we are just making a variable substitution for the Yang-Mills Equation.
\end{thebibliography}
\end{document}